\documentclass[12pt]{amsart}
\usepackage{amssymb, amsmath, amsfonts, amsthm}
\usepackage[utf8]{inputenc}
\usepackage{graphicx}
\usepackage[colorlinks=true, citecolor=blue, bookmarks=false]{hyperref}

\newcommand{\be}{{\bf e}}
\newcommand{\bu}{{\bf u}}
\newcommand{\bx}{{\bf x}}
\newcommand{\Sa}{\mathcal{S}_{\text{a}}}
\newcommand{\Sr}{\mathcal{S}_{\text{r}}}
\newcommand{\abs}[1]{\lvert #1 \rvert}
\newcommand{\ind}{\operatorname{ind}}
\newcommand{\sign}{\operatorname{sign}}

\usepackage{xcolor}
\newtheorem{theorem}{Theorem}

\newtheorem{remark}{Remark}
\newtheorem{lemma}{Lemma}

\title[Interior Structural Bifurcation  with Symmetry]{Interior Structural Bifurcation of 2D Symmetric Incompressible Flows}

\begin{document}

\author[Bozkurt]{Deniz Bozkurt}
\email{deniz.bozkurt@agu.edu.tr}
\address[DB]{Department of Mathematics, Erciyes University, 38039 Kayseri, Turkey \\
}

\author[Deliceoğlu]{Ali Deliceoğlu}
\email{adelice@erciyes.edu.tr}
\address[AD]{Department of Mathematics, Erciyes University, 38039 Kayseri, Turkey \\
}

\author[Şengül]{Taylan Şengül}
\email{taylan.sengul@marmara.edu.tr}
\address[TS]{Department of Mathematics, Marmara University, 34722 Istanbul, Turkey \\
}

\begin{abstract}
	The structural bifurcation  of a 2D divergence free vector field $\mathbf{u}(\cdot, t)$ when $\mathbf{u}(\cdot, t_0)$ has an interior isolated singular point $\mathbf{x}_0$ of zero index has been studied by Ma and Wang \cite{MW04}.
	Although in the class of divergence free fields which undergo a local bifurcation around a singular point, the ones with index zero singular points are generic, this class excludes some important families of symmetric flows.
	In particular, when $\mathbf{u}(\cdot, t_0)$ is anti-symmetric with respect to $\mathbf{x}_0$, or symmetric with respect to the axis located on $\mathbf{x}_0$ and normal to the unique eigendirection of the Jacobian $D\mathbf{u}(\cdot, t_0)$, the vector field must have index 1 or -1 at the singular point.
	Thus we study the structural bifurcation when $\mathbf{u}(\cdot, t_0)$ has an interior isolated singular point $\mathbf{x}_0$ with index -1, 1.
	In particular we show that if such a vector field with its acceleration at $t_0$ both satisfy aforementioned symmetries then generically the flow will undergo a local bifurcation.
  Under these generic conditions, we rigorously prove the existence of flow patterns such as pairs of co-rotating vortices and double saddle connections.
  We also present numerical evidence of Stokes flow in a rectangular and cylindrical cavity showing that the bifurcation scenarios we present are indeed realizable.
\end{abstract}

\keywords{Flow structures, structural stability, divergence-free vector
field and bifurcation.}

\maketitle

\section{ Introduction}
The main objective of this study is to classify the local bifurcations of incompressible 2D flows subject to certain symmetry constraints.
A remarkable result of the geometric theory of incompressible 2D flows developed by Ma and Wang \cite{MW05book} gives necessary and sufficient conditions for the structural stability of 2D incompressible flows \cite{MW02} and is an extension of the classical structural stability theorem of Peixoto \cite{Peixoto}.
Namely, a 2D divergence free vector field is structurally stable if and only if it satisfies these three properties: 1) $\bu$ is regular, 2) interior saddle points of $\bu$ are self-connected, 3) each saddle point on the boundary is connected to a saddle point on the same component of the boundary.
However this is a global result and is not applicable when only local information about the vector field is available.

The same research program also produced results on the local bifurcations around isolated singular points of 2D incompressible flows both near the boundaries (Ghil, Ma and Wang in \cite{GMW01}) and away from boundaries (Ma and Wang \cite{MW04}).
For this, consider the Taylor expansion of a 2D divergence-free vector field $\bu(\cdot, t)$ at $t=t_0$,
\[
	\bu (x, t) = \bu^0(x) + \bu^1(x) (t-t_0) + o(\abs{t-t_0})
\]
where
\[
	\bu^0(x) = \bu(x, t_0), \quad \bu^1(x) = \frac{\partial \bu}{\partial t} (x, t_0).
\]
Assume that $\bu^0$ has an interior ($\bx_0 \in \mathring{M}$) singular point ($\bu^0(\bx_0) = 0$) which is simple ($D\bu^0(\bx_0) \ne 0$) degenerate ($\det D \bu^0(\bx_0) = 0$), and isolated.
Let the unit vector $\be_1$ point in the unique eigendirection of $D\bu^0(\bx_0)$ and $\be_2$ be normal to it.
Ma and Wang \cite{MW04} proved that if $\bu^0$ has index zero at $\bx_0$ (i.e. $\bx_0$ is a degenerate cusp of $\bu^0$) and if the acceleration $\bu^1(\bx_0)$ does not vanish in the $\be_2$ direction, then a structural bifurcation occurs at time $t_0$.

This result is conclusive when there is no symmetry present since vector fields having singular points with zero index are generic in the class of all 2D divergence free vectors that undergo a local bifurcation.
However there are certain classes of symmetric flows which do not allow such singular points.
Most notably, when the flow is anti-symmetric with respect to the singular point (see equation~(\ref{anti sym})), or when the flow is axisymmetric with respect to $\be_2$ direction (see equation~(\ref{ref sym})).

Thus the main purpose of this paper is to extend the results of Ma-Wang \cite{MW04} to flows under such symmetry constraints.
We show that when such symmetries are present, the vector field $\bu(\cdot, t_0)$ has generically index -1 or 1 at $\bx_0$.
Suppose that $\bu^0$ has $\bx_0$ as a simple isolated interior degenerate singular point with index -1 (a saddle) or 1 (a center).
In this case we show that a generic perturbation does not give rise to a local bifurcation.
In particular, if the acceleration $\bu^1(\bx_0)$ does not vanish in the $\be_2$ direction, no local bifurcations will occur.
Thus we find sufficient conditions on the acceleration field for the existence of a local bifurcation.
These sufficient conditions reduce to the condition
\[
	\frac{\partial \bu^1 \cdot \be_2}{\partial \be_1} \mid_{\bx_0} \ne 0,
\]
when the acceleration itself satisfies the anti symmetry condition \eqref{anti sym}.

The bifurcation scenarios we obtain in \autoref{Thm: Bifurcation} are as follows.
In the case when $\bx_0$ is a saddle of $\bu^0$, the bifurcation occurs via the separation of the saddle into two saddles and a center, see \autoref{fig4}.
In the case when $\bx_0$ is a center of $\bu^0$, the bifurcation occurs via the separation of the center into two centers and a saddle (a figure eight), see \autoref{fig5}. We discuss the genericity of these two bifurcation scenarios for symmetric flows in \autoref{Theorem Genericity}.

We also would like to remark on differences between the structural stability and bifurcation for 2-D incompressible flows with symmetry which was previously studied by Hsia et al. \cite{hsia}.
The symmetry considered in their paper differs from the current work in the sense that in their work, the flow must be symmetric in the whole domain while we allow flows locally symmetric around a singular point.
Moreover, that paper is on the global bifurcations while our results are local.

Another very successful method to study local bifurcations is the consideration of the streamline topology near an interior point.
Therefore, there are many studies which examine streamline patterns and their bifurcations in two-dimensional incompressible flows near or away from the boundaries; see among many others (\cite{Andronov}, \cite{Perry}, \cite{Bakker91}) for a detailed discussion.
Brons and Hartnack \cite{brons1} was the first to investigate streamline patterns and their bifurcations near simple-degenerate critical point from a topological view.
In this approach, the Taylor expansion of the stream function is considered to obtain local information about a velocity field.
By considering the coefficients in the Taylor series as bifurcation parameters, a series of non-linear coordinate transformation is applied to simplify the stream function to obtain the normal form, i.e. the simplest possible higher-order terms near the interior point.
This approach was used to analyze a variety of specific steady flows, for example, a flow near a fixed wall \cite{hartnack}, slip flows (\cite{brons3}) a flow close to an axisymmetric flow \cite{brons4} and vortex breakdown (\cite{brons5}, \cite{brons6}).
The investigation of streamline topology and their bifurcation near a non-simple degenerate critical point close to a stationary wall and away from the boundaries were investigated by Deliceoglu and Gurcan (\cite{delice1}, \cite{delice2}, \cite{delice3}).

We would like to remark on the differences between the streamline topology approach and the topological index approach which we use in this study.
The topological index approach has the advantage that it allows us to consider flows that are not necessarily steady.
Hence it is possible to identify the role of acceleration field on the local bifurcation.
By carrying the homotopy invariance of the index, we can easily find the normal form of the stream function.

On the other hand, using normal form theory, it is easier to consider bifurcations of higher codimensions.
Moreover the role of the vorticity transport equation in steady flows can be determined by using a stream function obtained by a canonical transformation via a generating function.

The paper is outlined as follows. We give the setup of the problem in Section 2 and list our main results in Section 3.
The proofs of the main theorems is given in Section 4.
In Section 5, we demonstrate numerical evidence showing the validity of our main theorem in an application.
Finally Section 6 discusses the conclusions of this work.

\section{Setup of the Problem}
Let $M$  be a closed and bounded domain with $C^r$ ($r \ge 1$) boundary $\partial M$.
Let $TM$ denote the tangent bundle of $M$ and $C^r(TM)$ denote the $r$ times continuously differentiable vector fields on $M$.
Let
\[
  D^r(TM) = \{ v \in C^r(TM) \mid v \cdot n \mid_{\partial M} = 0, \, \text{div} \bu = 0 \},
\]
where $n$ is the outer normal on $\partial M$.

Now consider $\bu \in C^1 ([0, T], D^r(TM))$, i.e. a vector field in $D^r(TM)$ parametrized by $t$.
We say $\bu(x, t)$ has a local structural bifurcation in a neighborhood $U \subset M$ of $\bx_0$ at $t_0$ if for any sufficiently close $t^{-}$ and $t^{+}$ to $t_0$ with $t^{-} < t_0 < t^{+}$,  $\bu(\cdot, t^{-})$ and $\bu(\cdot, t^{+})$ are not topologically equivalent, i.e. no homeomorphism can be found which maps orbits of $\bu(\cdot, t^{-})$ to orbits of $\bu(\cdot, t^{+})$ preserving the orientation.

Consider the Taylor expansion of $\bu \in C^1([0, T], D^r(TM))$ (where $r$ is as large as necessary) at $t=t_0$,
\begin{equation} \label{u taylor t=t0}
\bu (x, t) = \bu^0(x) + \bu^1(x) (t-t_0) + o(\abs{t-t_0}),
\end{equation}
where
\begin{equation} \label{u taylor t=t0 coefs}
\bu^0(x) = \bu(x, t_0), \quad \bu^1(x) = \frac{\partial \bu}{\partial t} (x, t_0).
\end{equation}
We will think of the acceleration field $\bu^1$ as a small perturbation of the steady state flow $\bu^0$ near $t_0$.
In this paper we will consider the local bifurcations of flows having an isolated interior singular point $\bx_0 \in \mathring{M}$ at $t= t_0$, i.e.
\begin{equation} \label{u0(x0)=0}
\bu^0(\bx_0) = 0.
\end{equation}
It is known (see Ma and Wang~\cite{GMW01}) that $\bx_0$ is connected to $2n$ ($n \in \mathbb{N} \cup \{0\}$) orbits of $\bu^0$, $n$ of the orbits are stable (meaning that the omega-limit set of those orbits are $\{\bx_0\}$) and the rest are unstable (meaning that the alpha-limit set of those orbits are $\{\bx_0\}$).
Moreover the stable and unstable orbits connected to $\bx_0$ alternate when a closed curve is traced around $\bx_0$.
Furthermore,
\[
  \ind(\bu^0, \bx_0) = 1-n,
\]
where $\ind(\bu^0, \bx_0)$ is the index of $\bu^0$ at $\bx_0$ defined via the Brouwer degree.

$\bx_0$ is called a degenerate singular point of $\bu^0$ if the Jacobian determinant $\det D \bu^0(\bx_0)$ vanishes and non-degenerate otherwise.
Since local structural bifurcations do not occur around non-degenerate interior singular points, one focuses on degenerate ones to study such bifurcations.
Hence we will assume
\begin{equation} \label{u0 degenerate}
\det D \bu^0 (\bx_0) = 0.
\end{equation}
Generically it suffices to consider the case of simple degeneracy
\begin{equation} \label{u0 simple}
D \bu^0 (\bx_0) \ne 0,
\end{equation}
that is the Jacobian matrix does not vanish completely at $\bx_0$.

If $\bx_0 \in \mathring{M}$ is an isolated singular point with Jacobian $D \bu^0(\bx_0) \ne 0$ then a characterization given by Ma and Wang (Lemma~3.1, \cite{MW04}) states that one of the following must hold:
\begin{enumerate}
  \item either $\ind(\bu^0, \bx_0) = 1$ and $\bx_0$ is a degenerate center,
  \item or, $\ind(\bu^0, \bx_0) = -1$ and $\bx_0$ is a degenerate saddle such that 4 orbits connected to $\bx_0$ are tangent to each other at $\bx_0$,
  \item or $\ind(\bu^0, \bx_0) = 0$ and $\bx_0$ is a cusp such there are two orbits connected to $\bx_0$ and the angle between these orbits is zero.
\end{enumerate}
Thus isolated simple degenerate singular points of $\bu^0$ must have index $-1$, $0$ or $1$.

Due to simple degeneracy of $\bu^0$ at $\bx_0$, see \eqref{u0 degenerate} and \eqref{u0 simple}, there exists unit vectors $\be_1$, $\be_2$ satisfying
\begin{equation}\label{Du0(x0) eigenvectors}
  D \bu^0(\bx_0) \be_1 = 0, \qquad D \bu^0(\bx_0) \be_2 = \alpha \be_1,
\end{equation}
for some constant $\alpha \ne 0$.

Now we consider an orthogonal coordinate system $(x, y)$ with origin at $\bx_0$
\begin{equation} \label{x0=0}
  \bx_0 = 0,
\end{equation}
and the $x$ and $y$ axes pointing in the direction of $\be_1$ and $\be_2$ respectively.
In view of \eqref{Du0(x0) eigenvectors}, we have
\begin{equation} \label{Du0}
  D \bu^0(0) =
  \begin{bmatrix}
      0 & \alpha \\
      0 & 0
  \end{bmatrix},
  \qquad
  \alpha \ne 0.
\end{equation}

% Todo: Explain these two
Now assume further that
\begin{equation} \label{def of k}
  \frac{\partial ^{m} (\bu^0 \cdot \be_1) }{\partial x^m}(\bx_0) =
  \begin{cases}
  0, & 1\leq m < k, \\
  \neq 0, & m=k,
  \end{cases}
\end{equation}
and
\begin{equation} \label{def of n}
  \frac{\partial ^{m} (\bu^0 \cdot \be_2) }{\partial x^m}(\bx_0) =
  \begin{cases}
  0, & 1\leq m < n, \\
  \neq 0, & m=n.
  \end{cases}
\end{equation}

Under the above conditions \eqref{u0(x0)=0}--\eqref{def of n}, the Taylor series expansion of $\bu^0$ at $\bx_0$ becomes
\begin{equation} \label{u0 taylor expansion}
\bu^0(x, y) =
  \begin{cases}
    \alpha y + \lambda x^k + f(x, y) &  \\
    \beta x^n - \lambda k y x^{k-1} + g(x, y)
  \end{cases},
\end{equation}
where $f(x, y) = O(\abs{x}^{k+1}) + yO(\abs{x}) + O(y^2)$ and $g(x, y) = O(\abs{x}^{n+1}) + y O(\abs{x}^k) + O(y^2)$
\begin{equation} \label{u0 alpha beta lambda}
  \alpha, \beta, \lambda \in \mathbb{R}, \qquad
  \alpha \ne 0, \qquad \beta \ne 0, \qquad \lambda \ne 0,
\end{equation}
\begin{equation} \label{u0 n k}
  n, k \in \mathbb{Z}, \qquad
  n \ge 2, \qquad
  k \ge 2.
\end{equation}

As stated before, the index of $\bu^0$ at $\bx_0$ must be one of -1, 0, 1.
The following connection between the index of $\bu^0$ at $\bx_0$ and the coefficients of its Taylor series expansion at $\bx=\bx_0$ is made in \cite{MW04}.

\begin{figure}[ht]
    \centering
    \includegraphics[width=0.9\linewidth]{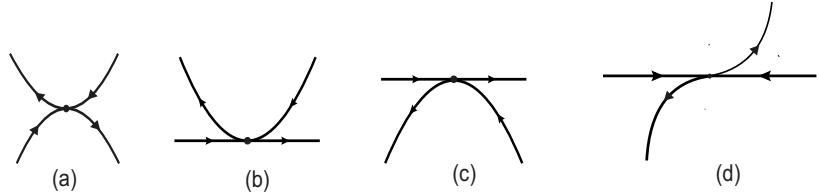}
    \centering
    \caption{Degenerate critical points with index $-1$.
    (a) $2k>n+1$, $n$ is odd, $\alpha \beta > 0$,
    (b) $n$ is odd, $k$ is even, $\alpha \lambda < 0$,
    (c) $n$ is odd, $k$ is even, $\alpha \lambda > 0$,
    (d) $n$ is odd, $k$ is odd.
    }
    \label{ty2}
\end{figure}

\begin{lemma}[Ma and Wang~\cite{MW04} Lemma~3.3] \label{index taylor coefficients lemma}
Consider the following complementary assumptions.
\begin{enumerate}
  \item[(S1)] $2k>n+1$, $n$ is even.
  \item[(S2)] $2k>n+1$, $n$ is odd, $\alpha \beta > 0$.
  \item[(S3)] $2k>n+1$, $n$ is odd, $\alpha \beta < 0$.
  \item[(S4)] $2k=n+1$, $\lambda^2 k + \alpha \beta > 0$.
  \item[(S5)] $2k=n+1$, $\lambda^2 k + \alpha \beta = 0$.
  \item[(S6)] $2k=n+1$, $\lambda^2 k + \alpha \beta < 0$.
  \item[(S7)] $2k<n+1$.
\end{enumerate}

The index of $\bu^0$ given by \eqref{u0 taylor expansion}--\eqref{u0 n k} at $\bx_0=0$ is
\begin{equation} \label{index fomula}
  \ind(\bu^0, \bx_0) =
  \begin{cases}
    0, & \text{if (S1) holds}, \\
   -1, & \text{if (S2), (S4) or (S7) holds}, \\
    1, & \text{if (S3) or (S6) holds}.
  \end{cases}
\end{equation}

\end{lemma}
Some remarks are in order:
\begin{enumerate}
  \item Since $\alpha \beta \ne 0$ by \eqref{u0 alpha beta lambda}, the assumptions given in \autoref{index taylor coefficients lemma} are indeed complementary.

  \item Note that under the assumption (S2) or (S7), the index of $\bu^0$ at $\bx_0$ is $-1$ and $\bu^0$ looks as shown in \autoref{ty2}.
  Under the assumption (S7), if $n$ is odd, $k$ is even and $\alpha \lambda <0$  (respectively, $\alpha \lambda >0$), then the flow pattern looks as shown in \autoref{ty2}(b) (respectively, as shown in \autoref{ty2}(c)).
  If both $n$ and $k$ are odd, the streamline pattern is as shown as in \autoref{ty2}(d).
  While the flow structures in \autoref{ty2} are topologically equivalent, they are geometrically different.
  We note that the degenerate critical points in \autoref{ty2}(b-d) were observed by Bakker \cite{Bakker91} and Hartnack \cite{hartnack} near a fixed wall.
  In this study, they appear away from the boundaries.
  \item Under the assumption (S5), the index of $\bu^0$ can not be determined from the Taylor expansion \eqref{u0 taylor expansion}--\eqref{u0 n k} alone and the higher order terms are required.
  In fact, in this case the truncated vector field (i.e. $\bu^0$ without the higher order terms) has a set of zeros on the curve $y= -\frac{\lambda}{\alpha} x^k$ and thus $\bx_0$ is not an isolated zero.
  \item We would like to point out the index formula \eqref{index fomula} is not given correctly in Ma and Wang~\cite{MW04} where it is stated that the index is always $-1$ if $\lambda^2 k + \alpha \beta \ne 0$, see (S4) and (S6) cases.
  We fix this issue by carrying out the proof of \autoref{index taylor coefficients lemma} under the assumptions (S4) and (S6) in Section~\ref{Section Proofs}.
\end{enumerate}

Now, consider $\bu^1$ in \eqref{u taylor t=t0 coefs} has the Taylor expansion at $\bx = \bx_0$ given by
\begin{equation}\label{bu1}
  \bu^1(x, y) =
  \begin{cases}
    \lambda_1 + O(\abs{\bx}), &  \\
    \lambda_0 + \lambda_2 x + \lambda_3 y + O(\abs{\bx}^2). &
  \end{cases}
\end{equation}

A genericity argument given in Ma and Wang~\cite{MW04} states that one only needs to study local bifurcations under the assumption (S1), i.e. when $\bu^0$ has $\bx_0$ as a simple degenerate cusp.
In that case, they prove the following theorem.

\begin{theorem} [Ma and Wang~\cite{MW04} Theorem~4.5] \label{Thm: Ma Wang bif}
Assume (S1) holds so that $\ind(\bu^0, \bx_0) = 0$ by \autoref{index taylor coefficients lemma} and $\lambda_0 = \bu^1(\bx_0) \cdot \be_2 \ne 0$.
Then under the assumptions \eqref{u taylor t=t0}--\eqref{def of n}, the vector field $\bu$ has a bifurcation in its local structure at $(\bx_0, t_0)$.
Moreover, for all $t$ sufficiently close to $t_0$, $\bu(\bx, t)$ has no singular points for any $t<t_0$ (resp. $t>t_0$) and exactly two non-degenerate singular points, one saddle and one center for all $t>t_0$ (resp. $t<t_0$).
\end{theorem}
% Todo: give picture of this theorem

However, symmetries considered in this paper destroy the possibility of local bifurcations around cusps.
Hence, one has to study local bifurcations around saddles and centers.
The situation can be compared to the occurrence of pitchfork bifurcations when the symmetry of the problem does not allow transcritical bifurcations.

To illustrate our point, we consider two classes of symmetric flows $\bu^0$ which do not fit into the picture of \autoref{Thm: Ma Wang bif}.
That is (S1) assumption can not be satisfied.

A vector field $\bu = (u, v)$ has reflectional symmetry about the y-axis if,
\begin{equation} \label{ref sym}
  u(x, y) = u (-x, y), \qquad v(x, y) = -v(-x, y),
\end{equation}
and has anti-symmetry with respect to the origin if,
\begin{equation} \label{anti sym}
  \bu(-\bx) = - \bu(\bx).
\end{equation}

When the vector field $\bu^0 = (u^0, v^0)$ given by \eqref{u0 taylor expansion} has reflectional symmetry about the $y$-axis, it is evident that $n$ is odd and $k$ is even. On the other hand, when it has anti-symmetry with respect to the origin, both $n$ and $k$ are odd.

Thus the assumption (S1) for $\bu^0$ which although is generic can not satisfy the above mentioned symmetries.

\section{Main Results}
Our main results are the following two theorems which are complementary to \autoref{Thm: Ma Wang bif}.

Our first result shows that a generic perturbation of the steady state $\bu^0$ will not cause a local bifurcation.
\begin{theorem} \label{Thm: No bifurcation}
Suppose one of the conditions (S2), (S3), (S4), (S6), (S7) holds so that $\ind(\bu^0, \bx_0) = \pm 1$ by \autoref{index taylor coefficients lemma}.
If $\lambda_0 = \bu^1(\bx_0) \cdot \be_2 \ne 0$ then under the assumptions \eqref{u taylor t=t0}--\eqref{def of n}, the vector field $\bu$ has no local bifurcation at $(\bx_0, t_0)$.
\end{theorem}
Under the assumptions of \autoref{Thm: No bifurcation}, the topological structure of the flow does not change as $t$ crosses $t_0$.
Merely, there is a unique singular point with index $\ind(\bu^0, \bx_0)$ of the perturbed flow near $\bx_0$ which becomes degenerate at $t=t_0$ and is non-degenerate when $t \ne t_0$ is sufficiently close to $t_0$.

Thus one needs to consider another class of perturbations $\bu^1$ which will give rise to a local bifurcation.
We address this issue in the next theorem.
\begin{theorem} \label{Thm: Bifurcation}
  Suppose one of the conditions (S2), (S3), (S4), (S6), (S7) holds so that $\ind(\bu^0, \bx_0) = \pm 1$ by \autoref{index taylor coefficients lemma}.

  In addition to $\lambda_0 =0$, assume also the following conditions on $\bu^1$ given by \eqref{bu1}.
  \begin{equation} \label{condition genericity}
    \begin{aligned}
    & 2 \lambda \lambda_1 + \alpha \lambda_2 \ne 0 && \text{if } k=2, \\
    & \lambda_2 = \frac{\partial \bu^1 \cdot \be_2}{\partial x} \mid_{\bx_0} \ne 0 && \text{if } k>2.
    \end{aligned}
  \end{equation}
  Then under the assumptions \eqref{u taylor t=t0}--\eqref{def of n}, $\bu$ has a bifurcation in its local structure at $(\bx_0, t_0)$ and the following statements hold true.
  \begin{enumerate}
    \item If $\ind(\bu^0, \bx_0) = -1$, then for all $t$ sufficiently close to $t_0$, $\bu(\bx, t)$ has exactly one non-degenerate saddle point for all $t<t_0$ (resp. $t>t_0$) and exactly three non-degenerate singular points, two saddles and a center for all $t>t_0$ (resp. $t<t_0$).
    The topological structure of the bifurcation is as shown in \autoref{fig4}.
    \begin{figure}[ht]
    \centering
    \includegraphics[width=0.5\linewidth]{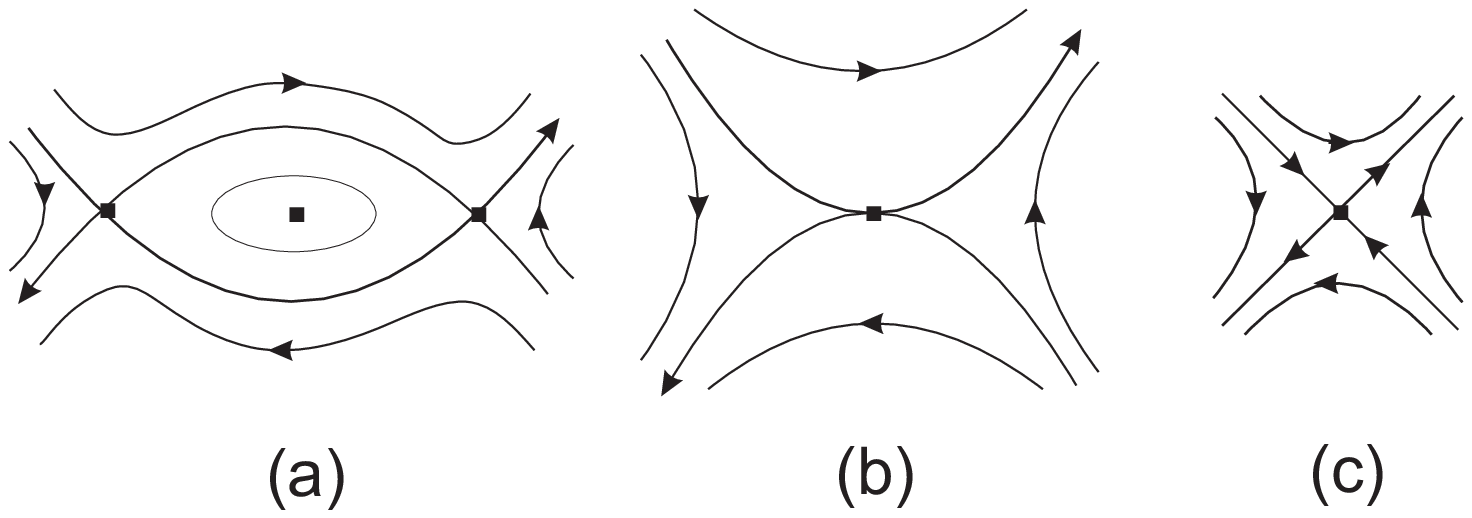}
    \centering
    \caption{Structural bifurcation diagram for $\ind(\bu^0, \bx_0) = -1$ case in \autoref{Thm: Bifurcation}:
    (a) $t=t_0+\epsilon$, (b) $t=t_0$, (c) $t= t_0 - \epsilon$.
    \label{fig4} }

    \end{figure}

    \item If $\ind(\bu^0, \bx_0) = 1$, then for all $t$ sufficiently close to $t_0$, $\bu(\bx, t)$ has exactly one non-degenerate center point for all $t<t_0$ (resp. $t>t_0$) and exactly three non-degenerate singular points, two centers and a saddle for all $t>t_0$ (resp. $t<t_0$).
    The topological structure of the bifurcation is as shown in \autoref{fig5}.
    \begin{figure}[ht]
    \centering
    \includegraphics[width=0.5\linewidth]{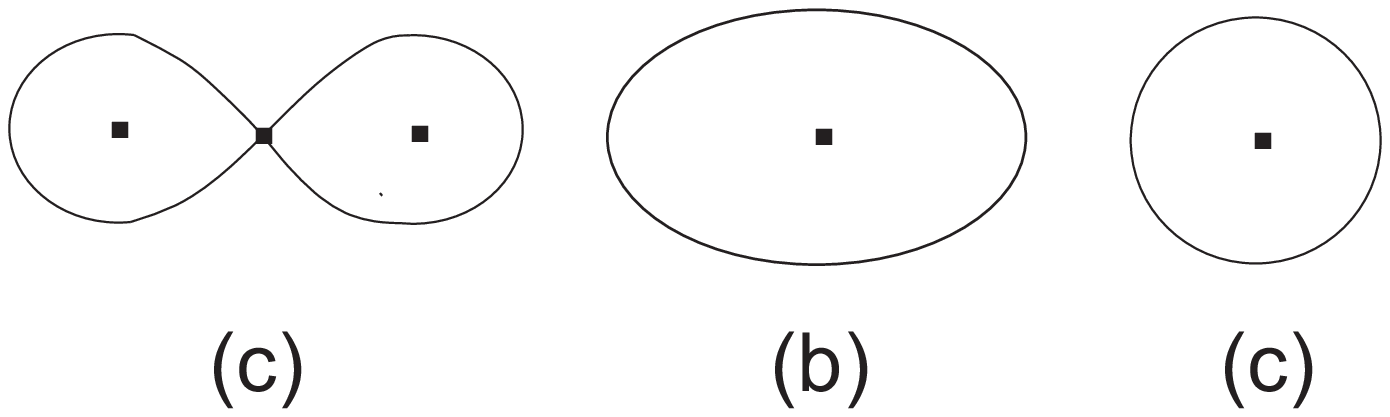}
    \centering
    \caption{Structural bifurcation diagram for $\ind(\bu^0, \bx_0) = 1$ case in \autoref{Thm: Bifurcation}:
    (a) $t=t_0+\epsilon$, (b) $t=t_0$, (c) $t= t_0 - \epsilon$.
    \label{fig5}
  }
    \end{figure}

  \end{enumerate}
\end{theorem}

\begin{remark} \label{Remark: symmetric u1}
  If the acceleration field $\bu^1$ is subject to the symmetry constraint \eqref{anti sym}, then $\lambda_1 = 0$ is automatically satisfied and the assumption \eqref{condition genericity} reduces to $\lambda_2 \ne 0$ which is equivalent to the condition
  \begin{equation*}
    \frac{\partial \bu^1 \cdot \be_2}{\partial x} \mid_{\bx_0} \ne 0.
  \end{equation*}
\end{remark}

\begin{remark}
  As stated in the introduction, a 2D divergence-free vector fields turn unstable if a connection between two saddle points, see Ma and Wang \cite{MW05book}.
  This is in contrast with the bifurcation scenario given in \autoref{fig4} and deserves an explanation.
  In \cite{MW05book}, such an instability is proven by a saddle-breaking technique which shows that any small (non-symmetric) perturbation near a saddle point may lead to the break-down of the saddle connection.
  However, as discussed by Hsia, Liu and Wang in \cite{hsia}, in the case of symmetric flows such connections are stable.
\end{remark}

\subsection*{Genericity of structural bifurcation in symmetric flows}
Now we discuss the genericity of the structural bifurcation given by \autoref{Thm: Bifurcation} for symmetric flows.

First consider the space of anti-symmetric vector fields
\[
  \Sa = \left\{
  \begin{aligned}
  & \bu \in C^1([0, T], D^r(TM)),
  \bu(-\bx) = - \bu(\bx), \\
  & \bu^0(\bx_0) =0, \det D\bu^0(\bx_0)=0, \, \bu^0 = \bu(\cdot, t_0)
  \end{aligned}
  \right\}.
\]
which contains all anti-symmetric smooth 2D divergence-free vector fields that undergo a local bifurcation at $(\bx_0, t_0)$.

In a similar fashion, consider the space of vector fields symmetric with respect to the y-axis that undergo a local bifurcation
\[
  \Sr = \left\{
  \begin{aligned}
  & \bu=(u, v) \in C^1([0, T], D^r(TM)),
  (u, v)(x, y) = (u, -v)(-x, y)
  \\
  & \bu^0(\bx_0) =0, \det D\bu^0(\bx_0)=0, \, \bu^0 = \bu(\cdot, t_0)
  \end{aligned}
  \right\}.
\]

\begin{theorem} \label{Theorem Genericity}
  There is an open and dense subset $\widetilde{\Sa}$ of $\Sa$ and an open and dense subset $\widetilde{\Sr}$ of $\Sr$ such that one of the local bifurcation scenarios stated in \autoref{Thm: Bifurcation} must hold for any $\bu \in \widetilde{\Sr}$ and for any $\bu \in \widetilde{\Sr}$.
\end{theorem}

\begin{remark}
  The result given in \autoref{Theorem Genericity} is valid even if one considers that instead of the whole flow $\bu$ only $\bu^0$ and $\bu^1$ are symmetric.
\end{remark}

\begin{remark}
  When $\ind(\bu^0, \bx_0) = -1$, the local bifurcation diagrams for symmetric flows become as given in \autoref{fig6} and \autoref{fig7}.

  \begin{figure}[ht]
  \centering
  \includegraphics[width=0.8\linewidth]{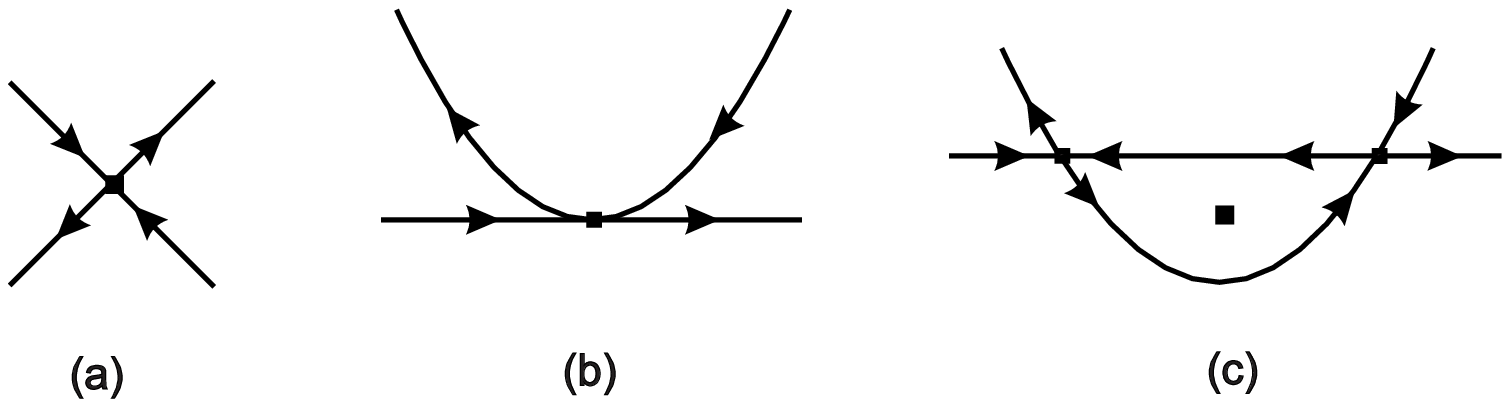}
  \centering
  \caption{ Structural bifurcation diagram for flows with reflectional symmetry \eqref{ref sym} in the $\ind(\bu^0, \bx_0) = -1$ case. (a) $t=t_{0}-\epsilon$, (b) $t=t_{0}$, (c) $t=t_{0}+\epsilon $ }
  \label{fig6}
  \end{figure}

  \begin{figure}[ht]
  \centering
  \includegraphics[width=0.8\linewidth]{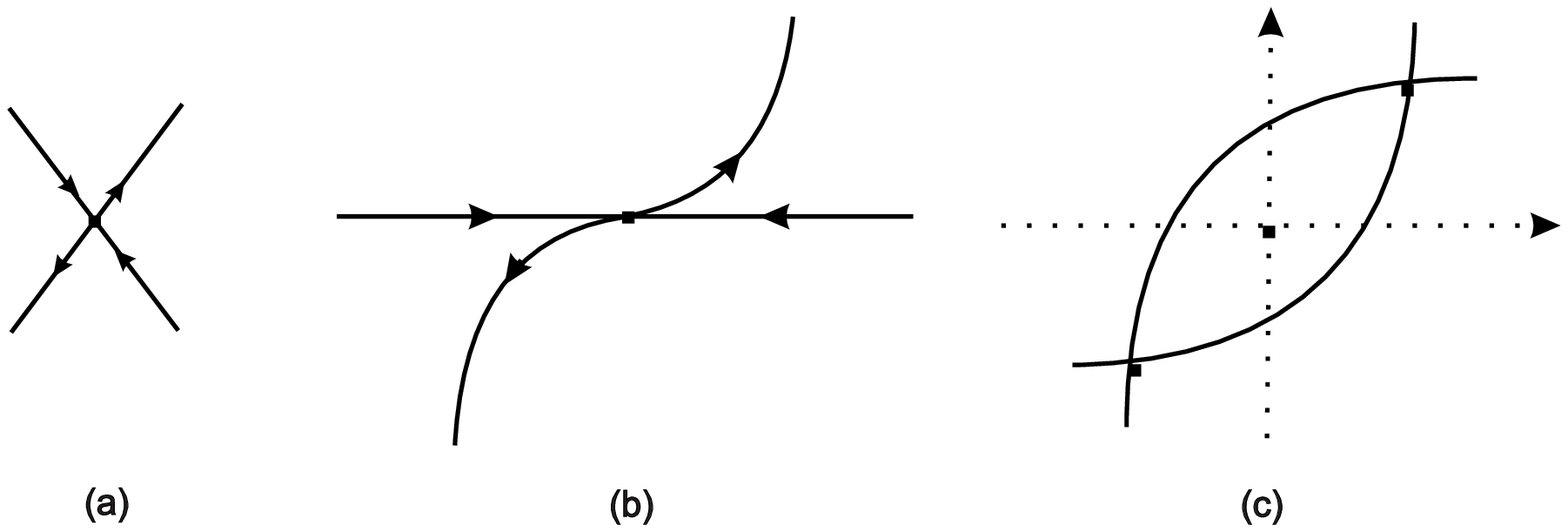}
  \centering
  \caption{ Structural bifurcation diagram for flows with anti symmetry \eqref{anti sym} in the $\ind(\bu^0, \bx_0) = -1$ case. (a) $t=t_{0}-\epsilon$, (b) $t=t_{0}$, (c) $t=t_{0}+\epsilon $ }
  \label{fig7}
  \end{figure}
\end{remark}

\section{Proofs} \label{Section Proofs}

\subsection*{Proof of \autoref{index taylor coefficients lemma} under the assumptions (S4) and (S6)}

Assume $2k = n+1$.
Let $\epsilon >0$ be sufficiently small and consider the following perturbation $\bu^0_{\epsilon}$ of $\bu^0$ given in \eqref{u0 taylor expansion}.
\begin{equation*}
\bu^0_{\epsilon }(\bx) = \bu^0 + (0, -\epsilon)^T.
\end{equation*}

In a small neighborhood of $(\bx) =0$, the singular points of $\bu^0_{\epsilon }$ satisfy the equations
\begin{equation} \label{3.10}
  y = - \frac{\lambda x^{k}}{\alpha} + O(\abs{x}^{k+1}).
\end{equation}
\begin{equation} \label{3.11}
  \beta x^n + \frac{1}{\alpha} k \lambda^2 x^{2k-1} = \epsilon + O (\abs{x}^{2k}).
\end{equation}
When $2k-1 = n$ and $\alpha \beta \ne k \lambda^2$, $k \ge 2$, the only solution of \eqref{3.10} and \eqref{3.11} is
\[
  \bx_{\epsilon} \sim \left( C \epsilon^{1/(2k-1)}, -\frac{\lambda}{\alpha} C^k \epsilon^{k/(2k-1)} \right),
\]
where $C = \alpha (\alpha \beta + k \lambda^2)^{-1}$.

Now it is easy to check that
\[
  \det D \bu^0_{\epsilon} (\bx_{\epsilon}) \sim -(C \epsilon)^{(2k-2)/(2k-1)} (2k-1) (\lambda^2 k + \alpha \beta).
\]
and
\[
  \sign \det D \bu^0_{\epsilon} (\bx_{\epsilon}) = - \sign (\lambda^2 k + \alpha \beta),
\]
so that
\[
  \ind(\bu^0_{\epsilon}, \bx_{\epsilon}) = -1.
\]

Now,
\[
  \ind(\bu^0, \bx_0) = \ind(\bu^0_{\epsilon}, \bx_{\epsilon}) = -1
\]
follows from the invariance of index under small perturbations.

\subsection*{Proof of \autoref{Thm: No bifurcation}.}

Assume that $\lambda_0 \ne 0$ and consider the singular points of the vector field $\bu^0 - \epsilon \bu^1 = 0$ for sufficiently small $\abs{\epsilon}$.
\begin{equation} \label{u0- ep u1 = 0}
\begin{aligned}
& \alpha y + \lambda x^k = \epsilon \lambda_1 + O(\abs{x}^{k+1}) + yO(\abs{x}) + O(y^2), \\
& \beta x^n - \lambda k y x^{k-1}  = \epsilon \lambda_0 + O(\abs{x}^{n+1}) + y O(\abs{x}^k) + O(y^2)
\end{aligned}
\end{equation}
By the implicit function theorem, the first equation in \eqref{u0- ep u1 = 0} can be solved uniquely for $y$ in a small neighborhood of $(\bx, \epsilon) = (0, 0)$.
\begin{equation} \label{y = func(x, eps)}
  y(x, \epsilon) = -\frac{1}{\alpha} \left( \lambda x^k - \epsilon \lambda_1 \right) + o(\abs{x}^k + \abs{\epsilon}).
\end{equation}
Using \eqref{y = func(x, eps)} and $\epsilon x^{k-1} = o(\abs{\epsilon})$ since $k \ge 2$, the second equation of \eqref{u0- ep u1 = 0} reduces to
\begin{equation} \label{lambda0 ne 0 case. x and eps relation}
  \beta x^n + \frac{\lambda^2 k}{\alpha} x^{2k-1}  = \epsilon \lambda_0 + o(\abs{x}^n + \abs{x}^{2k-1} + \abs{\epsilon}).
\end{equation}
The equation \eqref{lambda0 ne 0 case. x and eps relation} has the solution
\begin{equation} \label{eps=Cx^m}
  \epsilon = C x^m + o(\abs{x}^m), \quad C \ne 0, \quad
  m = \min\{ 2k-1,n \}
\end{equation}
which is the unique solution in a small neighborhood of $x=0$ and which can be inverted
\[
  x(\epsilon) = \frac{1}{C^{1/m}} \epsilon^{1/m} + o(\abs{\epsilon}^{1/m}), \quad
  m = \min\{ 2k-1,n \}
\]
in a small neighborhood of $\epsilon=0$ as both $n$ and $2k-1$, hence $m$ are odd.
In \eqref{eps=Cx^m}, $C \ne 0$ follows from the assumptions \eqref{u0 alpha beta lambda} and the fact that we have excluded the case
\[
  \lambda^2 k+ \alpha \beta \ne 0 \quad \text{when } 2k-1=n
\]
in the assumptions.
Let $y(\epsilon)$ be the solution determined by \eqref{y = func(x, eps)} corresponding to $x(\epsilon)$ and denote the solution
\[
  \bx(\epsilon) = (x(\epsilon), y(\epsilon)).
\]
Let us denote the Jacobian determinant of the perturbed field by
\[
  J = \det D (\bu^0 - \epsilon \bu^1).
\]

By \eqref{y = func(x, eps)} and \eqref{eps=Cx^m}, a straightforward computation yields
\[
  J(\bx(\epsilon)) = -\lambda^2 (k^2 + k (k-1)) x(\epsilon)^{2k-2} - n \alpha \beta x(\epsilon)^{n-1} + o(\abs{\epsilon}^{\frac{m-1}{m}}).
\]
From the above relation, we see that
\[
  \ind(\bu^0 - \epsilon \bu^1, \bx(\epsilon)) = J(\bx(\epsilon)) =
  \begin{cases}
    -\sign(\alpha \beta), & 2k > n+1 \\
    -\sign(\lambda^2 k + \alpha \beta), & 2k = n+1  \\
    -1, & 2k < n+1
  \end{cases}
\]
which equals to $\ind(\bu^0, \bx_0)$ by \autoref{index taylor coefficients lemma} for all $\epsilon$ sufficiently small.

Thus there is a unique solution of the perturbed field with the same index as $\ind(\bu^0, \bx_0)$.
Hence there is no bifurcation.
This finishes the proof.

\subsection*{Proof of \autoref{Thm: Bifurcation}}

Now suppose $\lambda_0 = 0$ and $\lambda_2 \ne 0$.
In this case, the equations for the singular points of the field $\bu^0 - \epsilon \bu^1 = 0$ are as follows.
\begin{equation} \label{lambda0=0 case u0- ep u1 = 0}
\begin{aligned}
 \alpha y + \lambda x^k = & \epsilon \lambda_1 + O(\abs{x}^{k+1}) + y O(\abs{x}) + O(y^2) + \epsilon O(\abs{\bx}), \\
 \beta x^n - \lambda k y x^{k-1}  = & \epsilon \lambda_2 x + \epsilon \lambda_3 y + O(\abs{x}^{n+1}) + y O(\abs{x}^k) + \\
& O(y^2) + \epsilon O(\abs{\bx}^2).
\end{aligned}
\end{equation}

Note that since the first equation of \eqref{lambda0=0 case u0- ep u1 = 0} is the same as the first equation of \eqref{u0- ep u1 = 0}, we still have the unique solution for $y$ in terms of $x$ and $\epsilon$ given by \eqref{y = func(x, eps)} in a small neighborhood of $(\bx, \epsilon) = (0, 0)$.
Plugging \eqref{y = func(x, eps)} into the second equation of \eqref{lambda0=0 case u0- ep u1 = 0}, and using
\begin{equation} \label{}
  \begin{aligned}
  & y O(\abs{x}^k) = O(\abs{x}^{2k}) + \epsilon O(\abs{x}^k), \\
  & O(y^2) = O(x^{2k}) + \epsilon O(\abs{x}^k) + O(\epsilon^2), \\
  & \epsilon y = \epsilon O(\abs{x}^k) + O(\epsilon^2),
  \end{aligned}
\end{equation}
yields
\begin{equation} \label{x ne 0 solutions}
  \begin{aligned}
  x & \left( \alpha \beta x^{n-1} + k \lambda^2 x^{2k-2} - \epsilon \left( \alpha \lambda_2 + k \lambda \lambda_1 x^{k-2} \right)  \right) = \\
  & O(\abs{x}^{n+1}) + y O(\abs{x}^{2k}) + \epsilon O(\abs{x}^2) + O(\epsilon^2). \\
  \end{aligned}
\end{equation}
As $\alpha \lambda_2 \ne 0$, the equations \eqref{lambda0=0 case u0- ep u1 = 0} have always the solution given by
\[
  \bx_0(\epsilon)= (x_0(\epsilon), y_0(\epsilon)) = (O(\abs{\epsilon}), \epsilon \lambda_1/ \alpha + o(\abs{\epsilon})) ,
\]
which can be obtained from the balance between the $\epsilon x$ term and the $O(\epsilon^2)$ term in \eqref{x ne 0 solutions}. For this, we look for a solution of \eqref{x ne 0 solutions} of the form $x_0 = z(\epsilon) \epsilon$, $z(\epsilon)=O(1)$ as $\epsilon\to 0$, and use the Implicit Function Theorem to deduce the existence of such a solution. The exact computation of the lowest order approximation of $x_0(\epsilon)$ requires the higher order terms in \eqref{x ne 0 solutions}, which is not required for what follows.
The vector field $\bu^0 - \epsilon \bu^1$ has the Jacobian
\[
  J(\bx_0(\epsilon)) = \alpha \epsilon \lambda_2 + o(\abs{\epsilon}) +
    \begin{cases}
    0  , &  k >2, \\
    2 \epsilon \lambda \lambda_1  , & k = 2.
    \end{cases}
\]
at the singular point $\bx_0(\epsilon)$.

Since both $n-1$ and $2k-2$ are both odd, there are two other solutions $\bx_{\pm}(\epsilon)$ of \eqref{lambda0=0 case u0- ep u1 = 0} determined by the solutions $x_{\pm}(\epsilon)$ of \eqref{x ne 0 solutions} near the origin which bifurcate on one side of $\epsilon$. To the lowest order approximation $x_{\pm}$ can be obtained from the truncated equation of \eqref{x ne 0 solutions}
\begin{equation} \label{x pm}
  \alpha \beta x^{n-1} + k \lambda^2 x^{2k-2} - \epsilon \left( \alpha \lambda_2 + k \lambda \lambda_1 x^{k-2} \right) = 0.
\end{equation}
Let $y_{\pm}(\epsilon)$ correspond to $x_{\pm}(\epsilon)$ via \eqref{y = func(x, eps)} and define
\[
  \bx_{\pm} (\epsilon) = (x_{\pm}(\epsilon), y_{\pm}(\epsilon)).
\]
The nature of these solutions depend on the parameters which we investigate in detail below.
In particular, we have to consider the cases $k=2$ and $k>2$ separately.

The perturbed field $\bu^0 - \epsilon \bu^1$ has one singular point on $\epsilon < 0$ (resp. $\epsilon > 0$) which is non-degenerate and the index of $\bu^0 - \epsilon \bu^1$ at that singular point is equal to $\ind(\bu^0, \bx_0)$.
On $\epsilon> 0$ (resp. $\epsilon < 0$) there are three non-degenerate singular points such that sum of the indexes of $\bu^0 - \epsilon \bu^1$ at these singular points equals $\ind(\bu^0, \bx_0)$.
This result is a direct consequence of the homotopy invariance of index sums in a small domain.
Hence there remains to describe the details of the bifurcation in each case by explicitly computing the bifurcated solutions and their Jacobians which we present below.
These details also verify the above claims in each case.

\subsection*{Case \texorpdfstring{$k=2$}.}
By the condition \eqref{condition genericity} we have $2 \lambda \lambda_1 + \alpha \lambda_2 \ne 0$.
Assume without loss of generality that $2 \lambda \lambda_1 + \alpha \lambda_2 > 0$.
The case $2 \lambda \lambda_1 + \alpha \lambda_2 < 0$ is dealt similarly.

Notice that since $n \ge 2$, (S2) and (S3) assumptions need not be checked when $k=2$.

\subsubsection*{Subcase $k=2$ and $n=3$.}
This subcase corresponds to either (S4) or the (S6) assumptions. In either case, we have $\alpha \beta + 2 \lambda^2 \ne 0$.
From \eqref{x pm},
\begin{equation} \label{eq: k=2, n=3}
  \begin{aligned}
  & x_{\pm}(\epsilon) = \pm \left( \frac{\left( 2 \lambda \lambda_1 + \alpha \lambda_2 \right) }{\alpha \beta + 2 \lambda^2} \epsilon \right)^{1/2} + o(\abs{\epsilon}^{1/2}), \\
  & J(\bx_{\pm}(\epsilon)) = -2 (2 \lambda \lambda_1 + \alpha \lambda_2)\epsilon + o(\abs{\epsilon}).
  \end{aligned}
\end{equation}

By \eqref{eq: k=2, n=3}, the bifurcated solutions are as shown in \autoref{tab:k=2,n=3}.

\begin{table}[th]
  \caption{The bifurcated solutions for $k=2$, $n=3$, $2 \lambda \lambda_1 + \alpha \lambda_2 > 0$.}
  \label{tab:k=2,n=3}
  \centering

  \begin{tabular}{|c|c|c|}
  \hline
  & $\epsilon<0$ & $\epsilon>0$ \\
  \hline
  $\alpha \beta + 2 \lambda^2 > 0$   & $\bx_{0}(\epsilon)$ is a saddle & $\bx_{\pm}(\epsilon)$ are saddles, $\bx_{0}(\epsilon)$ is a center\\
  \hline
  $\alpha \beta + 2 \lambda^2 < 0$   & $\bx_{0}(\epsilon)$ is a center & $\bx_{\pm}(\epsilon)$ are centers, $\bx_{0}(\epsilon)$ is a saddle\\
  \hline
  \end{tabular}
\end{table}

\subsubsection*{Subcase $k=2$ and $n>3$.}
This subcase corresponds to (S7) assumption.
Since by \eqref{u0 alpha beta lambda}, we have $\lambda \ne 0$ and
\begin{equation} \label{eq: k=2, n>3}
  \begin{aligned}
  & x_{\pm}(\epsilon) = \pm \left( \frac{\left( 2 \lambda \lambda_1 + \alpha \lambda_2 \right) }{2 \lambda^2} \epsilon \right)^{1/2}  + o(\abs{\epsilon}^{1/2}), \\
  & J(\bx_{\pm}(\epsilon)) = -2 (2 \lambda \lambda_1 + \alpha \lambda_2)\epsilon + o(\abs{\epsilon}).
  \end{aligned}
\end{equation}

By \eqref{eq: k=2, n>3}, the bifurcated solutions are as shown in \autoref{tab:k=2,n>3}.

\begin{table}[th]
  \caption{The bifurcated solutions for $k=2$, $n>3$, $ 2 \lambda \lambda_1 + \alpha \lambda_2 > 0$.}
  \label{tab:k=2,n>3}
  \centering

  \begin{tabular}{|c|c|c|}
  \hline
  $\epsilon<0$ & $\epsilon>0$ \\
  \hline
  $\bx_{0}(\epsilon)$ is a saddle & $\bx_{\pm}(\epsilon)$ are saddles, $\bx_{0}(\epsilon)$ is a center\\
  \hline
  \end{tabular}
\end{table}

\subsection*{Case \texorpdfstring{$k>2$}.}
In this case, by \eqref{condition genericity} we have $\lambda_2 \ne 0$.
Since $\alpha \ne 0$, without loss of generality $\alpha \lambda_2 > 0$.
The case $\alpha \lambda_2 < 0$ is dealt similarly.

\subsubsection*{Subcase $k>2$ and $2k<n+1$.}
This subcase corresponds to (S7) assumption.
Since $k \ne 0$ and $\lambda \ne 0$ by \eqref{u0 alpha beta lambda} and \eqref{u0 n k}, we have
\begin{equation} \label{eq: k>2, 2k<n+1}
  \begin{aligned}
  & x_{\pm}(\epsilon) = \pm  \left( \frac{\epsilon \alpha \lambda_2}{k \lambda^2 } \right)^{\frac{1}{2k-2}} + o(\abs{\epsilon}^{\frac{1}{2k-2}}), \\
  & J(\bx_{\pm}(\epsilon)) = -(2k-2) \epsilon \alpha \lambda_2+ o(\abs{\epsilon}).
  \end{aligned}
\end{equation}

By \eqref{eq: k>2, 2k<n+1}, the bifurcated solutions of are as shown in \autoref{tab:k>2,2k<n+1}.

\begin{table}[th]
  \caption{The bifurcated solutions for $k>2$, $2k<n+1$, $\alpha \lambda_2 >0$.}
  \label{tab:k>2,2k<n+1}
  \centering
  \begin{tabular}{|c|c|c|}
  \hline
  $\epsilon<0$ & $\epsilon>0$ \\
  \hline
  $\bx_{0}(\epsilon)$ is a saddle & $\bx_{\pm}(\epsilon)$ are saddles, $\bx_{0}(\epsilon)$ is a center\\
  \hline
  \end{tabular}
\end{table}

\subsubsection*{Subcase $k>2$ and $2k>n+1$.}
This subcase corresponds to (S2) and (S3) assumptions.
Since $\beta \ne 0$ by \eqref{u0 alpha beta lambda}, we have
\begin{equation} \label{eq: k>2, 2k>n+1}
  \begin{aligned}
  & x_{\pm}(\epsilon) = \pm  \left( \frac{\epsilon \lambda_2}{ \beta } \right)^{\frac{1}{n-1}} + o(\abs{\epsilon}^{\frac{1}{n-1}}), \\
  & J(\bx_{\pm}(\epsilon)) = - (n-1) \alpha \epsilon \lambda_2 + o(\abs{\epsilon}).
  \end{aligned}
\end{equation}

By \eqref{eq: k>2, 2k>n+1}, the bifurcated solutions are as shown in \autoref{tab:k>2,2k>n+1}.

\begin{table}[th]
  \caption{The bifurcated solutions for $k>2$, $2k>n+1$, $\alpha \lambda_2>0$.}
  \label{tab:k>2,2k>n+1}
  \centering

  \begin{tabular}{|c|c|c|}
  \hline
  & $\epsilon<0$ & $\epsilon>0$ \\
  \hline
  $\alpha \beta > 0$   & $\bx_{0}(\epsilon)$ is a saddle & $\bx_{\pm}(\epsilon)$ are saddles, $\bx_{0}(\epsilon)$ is a center\\
  \hline
  $\alpha \beta < 0$   & $\bx_{0}(\epsilon)$ is a center & $\bx_{\pm}(\epsilon)$ are centers, $\bx_{0}(\epsilon)$ is a saddle\\
  \hline
  \end{tabular}
\end{table}

\subsubsection*{Subcase $k>2$ and $2k=n+1$.}
This subcase corresponds to (S4) and (S6) assumptions. Under both assumptions we have $\alpha \beta + k \lambda^2 \ne 0$ and we have
\begin{equation} \label{eq: k>2, 2k=n+1}
  \begin{aligned}
  & x_{\pm}(\epsilon) = \pm  \left( \frac{\epsilon \alpha \lambda_2}{ \alpha \beta + k \lambda^2 } \right)^{\frac{1}{2k-2}} + o(\abs{\epsilon}^{\frac{1}{2k-2}}), \\
  & J(\bx_{\pm}(\epsilon)) = - (2k-2) \alpha \epsilon \lambda_2 + o(\abs{\epsilon}).
  \end{aligned}
\end{equation}
By \eqref{eq: k>2, 2k=n+1}, the bifurcated solutions are as shown in \autoref{tab:k>2,2k=n+1}.

\begin{table}[th]
  \caption{The bifurcated solutions for $k>2$, $2k=n+1$, $\alpha \lambda_2 > 0$.}
  \label{tab:k>2,2k=n+1}
  \centering

  \begin{tabular}{|c|c|c|}
  \hline
  & $\epsilon<0$ & $\epsilon>0$ \\
  \hline
  $ \alpha \beta + k \lambda^2 > 0$   & $\bx_{0}(\epsilon)$ is a saddle & $\bx_{\pm}(\epsilon)$ are saddles, $\bx_{0}(\epsilon)$ is a center\\
  \hline
  $ \alpha \beta + k \lambda^2 < 0$   & $\bx_{0}(\epsilon)$ is a center & $\bx_{\pm}(\epsilon)$ are centers, $\bx_{0}(\epsilon)$ is a saddle\\
  \hline
  \end{tabular}
\end{table}

\subsection*{Proof of \autoref{Theorem Genericity}}
Let
\[
  \widetilde{\Sa} =
  \left\{
    \begin{aligned}
    & \bu \in \Sa, \,
    D \bu^0(\bx_0) \ne 0,
    \text{\eqref{def of n} holds with $n=3$, } \\
    & \text{\eqref{def of k} holds with $k=3$, }
    \lambda^2 k + \alpha \beta \ne 0, \,\lambda_2 \ne 0
    \end{aligned}
  \right\}.
\]
It is easy to see that the space $\widetilde{\Sa}$ is open and dense in $\Sa$. Moreover for $\bu \in \widetilde{\Sa}$ the assumptions (S1) and (S5) of \autoref{index taylor coefficients lemma} can not hold.
Hence one of the assumptions (S2), (S3), (S4), (S6) or (S7) must hold.
Finally for $\bu \in \widetilde{\Sa}$, the condition \eqref{condition genericity} holds, see \autoref{Remark: symmetric u1}.
Now \autoref{Thm: Bifurcation} gives the desired result.

Similarly we define
\[
  \widetilde{\Sr} =
  \left\{
    \begin{aligned}
    & \bu \in \Sr, \,
    D \bu^0(\bx_0) \ne 0,
    \text{\eqref{def of n} holds with $n=3$, } \\
    & \text{\eqref{def of k} holds with $k=2$, }
    \lambda^2 k + \alpha \beta \ne 0, \,
    \eqref{condition genericity} \text{ holds}.
    \end{aligned}
  \right\}
\]
which is clearly an open and dense subset of $\mathcal{S}_r$. Once again the conditions of \autoref{Thm: Bifurcation} are satisfied. This finishes the proof.

\section{Numerical Evidence}
In this section we present numerical evidence showing that the bifurcation scenario given by \autoref{Thm: Bifurcation} is actually observed in certain flow scenarios.

\begin{figure}
\includegraphics{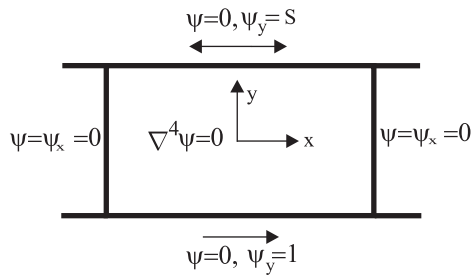}
\caption{Illustration of the dimensionless boundary value problem. }\label{22}
\end{figure}

\begin{figure}
\includegraphics[width=130mm]{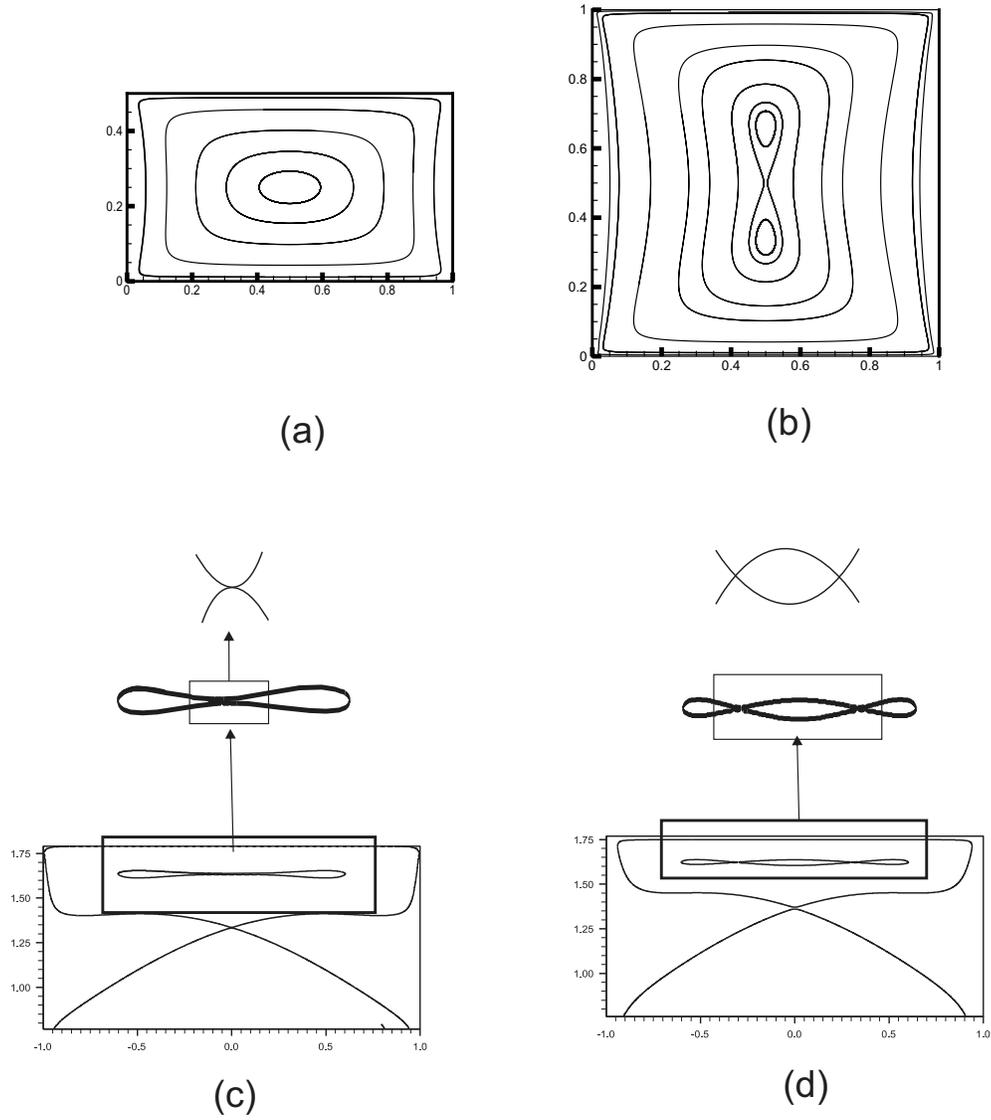}
\caption{Streamlines patterns a in rectangular cavities.  (a) $(A,S)=(0.5,-1)$, (b)
$(A,S)=(1,-1)$, (c) $(A,S)=(4.678,-0.0031)$, (d) $(A,S)=(4.627,-0.0031)$ }\label{fig22a}
\end{figure}

The first application we consider is the steady flow inside a double-lid-driven rectangular cavity.
A more detailed description of this application can be found in Gurcan and Deliceoglu (\cite{delice3}) and Gurcan (\cite{gurcan}).
Gurcan and Deliceoglu (\cite{delice3}) examined various sequences of flow transitions in the double-lid-driven cavity with the lids moving in opposite directions (reproduced in Figure \ref{22}).
In this problem, the stream function solution is obtained in a cavity with two control parameters: the cavity aspect ratio ($A$) and the speed ratio of the lids ($S$).
For $A=0.5$, a single eddy occupies the cavity as shown in Figure \ref{fig22a}(a).
As aspect ratio is increased, there is a pitchfork bifurcation at $A=0.931$ when the center becomes a saddle with two sub-eddies (see Figure \ref{fig22a}(b)).
Figures \ref{fig22a}(c)- (d) show an example of topological saddle point bifurcation.
Figure \ref{fig22a}(c) shows a separatrix with one saddle point, two sub-eddies with centers for the case $S=-0.0031$ and $A=4.678$. By decreasing $A$ to around $4.627$ a topological saddle point bifurcation appears above the cavity.
There are now two separatrices with ``treble eddy''. This flow structure is demonstrated in Figure \ref{fig22a}(d)). These flow separations are exactly as described by our main theorems.

\begin{figure}
\includegraphics[width=70mm]{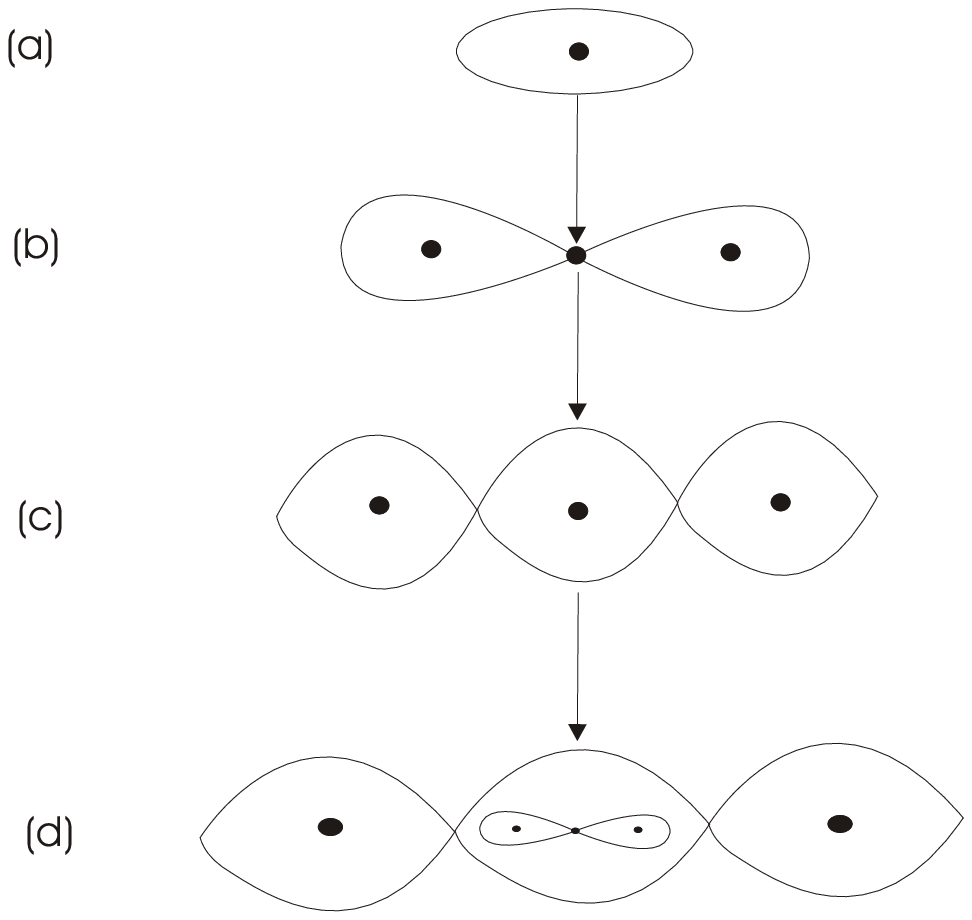}
\caption{Schematics showing flow structures obtained by Gaskell et al. \cite{gaskell} in a cylindrical cavity.  }\label{gaskell}
\end{figure}

Gaskell et al. \cite{gaskell} studied the Stokes flow in a half-filled annulus between rotating coaxial cylinders.
They investigated flow structures in a cylindrical cavity with two control parameters: the ratio $\overline{R}$ of the radii of the cylinders and the ratio $S$ of the peripheral speeds of the cylinders.
By variation of the radius ratio ($\overline{R}$), the flow patterns (a)-(d) in Figure \ref{gaskell} were obtained by Gaskell \cite{gaskell}.
The topological structure of the bifurcated flow is, once again, exactly as described in our main theorem.

\section{Conclusion}
In this work, we extend the results obtained in Ma and Wang~\cite{MW04} on the local bifurcations of flows to some non-generic cases which include certain symmetric flows.
In particular we study the local bifurcations at time $t_0$ of the flow $\bu$ when $\bu(\cdot, t_0)$ has a simple interior degenerate singular point of index -1 (a saddle) or 1 (a center).
We show that if the acceleration field at $t_0$ does not satisfy a certain genericity condition then there is no bifurcation.
Thus we give sufficient conditions on the acceleration field for which the flow will undergo a local bifurcation.
We also show that the two structural bifurcation scenarios we obtain are indeed generic for flows with certain symmetries.
Finally, we present numerical evidence of the Stokes flow in both a rectangular cavity and a cylindrical cavity showing that the bifurcation scenarios we prove are indeed realizable.

\end{document}